\documentclass[aps,prl,twocolumn,groupedaddress,preprintnumbers,nofootinbib,balancelastpage]{revtex4-1}

\usepackage{graphicx}
\usepackage{amssymb}
\usepackage{amsmath}
\usepackage{hyperref}
\usepackage{xspace}
\usepackage{paralist}
\usepackage{epsfig}
\usepackage{psfrag}
\usepackage{epstopdf}
\usepackage{color}
\usepackage{ragged2e}
\usepackage{float}
\usepackage{comment}

\begin{document}

\title{Enhancing charge ratio sensitivity to hadronization effects via jet selections on \textit{resolved SoftDrop splitting}}

\author{Liliana Apolin\'ario}
\email{liliana@lip.pt}
\affiliation{LIP, Av. Prof. Gama Pinto, 2, P-1649-003 Lisbon, Portugal}
\affiliation{Instituto Superior T\'{e}cnico (IST), Universidade de Lisboa, Av. Rovisco Pais 1, 1049-001, Lisbon, Portugal}

\author{Nuno Olavo Madureira}
\email{nuno.olavo@tecnico.ulisboa.pt}
\affiliation{LIP, Av. Prof. Gama Pinto, 2, P-1649-003 Lisbon, Portugal}
\affiliation{Instituto Superior T\'{e}cnico (IST), Universidade de Lisboa, Av. Rovisco Pais 1, 1049-001, Lisbon, Portugal}

\author{Raghav Kunnawalkam Elayavalli}
\email{raghav.ke@vanderbilt.edu}
\affiliation{Department of Physics and Astronomy, Vanderbilt University, Nashville, TN}

\date{\today}

\begin{abstract}
The study of Quantum Chromodynamics (QCD) at ultra-relativistic energies can be performed in a controlled environment through lepton-hadron deep inelastic scatterings. In such collisions, the high-energy partonic emissions that follow from the ejected hard partons are accurately described by perturbative QCD. However, the lower energy scales at which quarks and gluons experience colour confinement, i.e. hadronization mechanism, fall outside the validity regions for perturbative calculations, requiring phenomenological models tuned to data to describe it. As such, hadronization physics cannot be currently derived from first principles alone. Monte Carlo event generators are useful tools to describe these processes as they simulate both the perturbative and the non-perturbative interactions, with model-dependent energy scales that control parton dynamics. This work employs jets - experimental reconstructions of final-state particles likely to have a common partonic origin - to inspect this transition further. Although originally proposed to circumvent hadronization effects, we show that jets can be utilised as probes of non-perturbative phenomena via their substructure. The charge correlation ratio was recently shown to be sensitive to hadronization effects. Our work further improves this sensitivity to non-perturbative scales by introducing a new selection based on the relative placement of the \textit{resolved SoftDrop splitting} within the clustering tree, defined as the unclustering that resolves the jet's leading charged particles.
\end{abstract}

\maketitle

\section{I - Introduction}

The concept of jets and their substructure dates back to the first evidence of quarks and gluons from collimated sprays of hadrons in annihilation experiments of electrons and positrons~\cite{Riordan:1992hr, Ellis:2014rma, TASSO:1979zyf}. Since then, significant advances in fundamental QCD have been made due to copious production of jets in relativistic hadron-hadron colliders~\cite{Ali:2010tw}. Many of the state-of-the-art Monte-Carlo (MC) models~\cite{Bahr:2008pv,Sherpa:2019gpd,Sjostrand:2006za,Bierlich:2022pfr} that generate full collision events rely on both a perturbative Quantum Chromodynamics (pQCD) description of parton showers (including a resummation over all orders) and non-perturbative QCD (npQCD) implementation of hadronization and underlying event mechanisms. These parameterized implementations rely on fitting legacy $e+e-$ annihilation data to Deep Inelastic Scatterings (DIS) and, finally, recent data from hadron colliders~\cite{Skands:2010ak, Aguilar:2021sfa}. 

Their treatment of hadronization has to reflect the non-perturbative conversion of the coloured, outgoing final-state partons into colourless hadrons through a combination of analytical results and different QCD-based phenomenological models. There are two main classes of hadronization models - the Lund string model~\cite{Andersson:1983jt,Andersson:1983ia, Sjostrand:1984ic,Andersson:1997xwk, PhysRevD.46.2636}, implemented for instance in \textit{Pythia}~\cite{Bierlich:2022pfr}, and the cluster fragmentation approach~\cite{Amati:1979fg}, found in \textit{Herwig}~\cite{Bahr:2008pv} and \textit{Sherpa}~\cite{Gleisberg:2008ta} MC event generators. While the Lund string model depicts hadronization after the fragmentation of massless, one-dimensional strings stretching between colour-connected charges, the cluster hadronization framework is based on the colour ``preconfinement" property and prescribes the fragmentation of on-shell, colour singlet lumps of quarks and gluons - the clusters. For more information, we refer the reader for the references above.

Recent effort in the high-energy physics community has been in the area of developing novel experimental algorithms that translate a jet clustering tree to a theoretically motivated description of a parton shower~\cite{Dasgupta:2001sh, Dasgupta:2013ihk, Larkoski:2014wba, Mehtar-Tani:2019rrk}. Measurements from the Relativistic Heavy-Ion Collider (RHIC) and from the Large Hadron Collider (LHC) have shown that jet substructure observables, when calculated via grooming techniques such as SoftDrop (SD)~\cite{CMS:2017qlm, CMS:2018fof, ATLAS:2019mgf, ALICE:2019ykw, ATLAS:2020bbn, STAR:2020ejj, STAR:2021lvw, ALICE:2021njq, ALICE:2022hyz, ALICE:2022rdg}, allow one to fit pQCD calculations at the first hard splitting without the need for large hadronization corrections. Namely, the groomed momentum fraction ($z_g$), \textit{i.e.} the fractional transverse momentum of the softest subjet in a QCD jet, fits very well to the pQCD splitting functions when evaluated at the first SD splitting. For subsequent emissions along the higher energy (hard) branch of a jet shower~\cite{Dreyer:2018tjj}, the shape of the splitting function gradually changes from a steeply falling pQCD-like $1/z_g$ distribution to a flatter shape due to phase space restrictions. However, so as the impact of npQCD effects~\cite{Robotkova:2022jgn, KunnawalkamElayavalli:2022vys}. Previous works~\cite{Apolinario:2020uvt} demonstrated that SD groomed clustering trees significantly improve the correlation between the formation time of the parton shower emissions and the formation time of the first jet SD unclustering step, regardless of choice of jet clustering algorithm~\cite{Apolinario:2020uvt}. However, subsequent unclustering steps also observe their parton shower correlation worsen, although with a large dependence on the chosen algorithm.

The current state of SoftDrop substructure studies therefore requires the assessment of non-perturbative contributions to the jet evolution history, as they become more and more meaningful to the late-time jet dynamics. This calls for the exploration of experimentally robust jet substructure observables sensitive to hadronization physics. In~\cite{Chien:2021yol} the charge correlation ratio was proposed as a possible observable that is able to discriminate between the two aforementioned hadronization models: Lund string and cluster fragmentation. In this manuscript, we show that jet substructure selection can largely increase this sensitivity, allowing for more stringent tests to hadronization mechanisms. For that, this manuscript is organized as follows: in section II we present the details of event generation using \textit{Pythia} and \textit{Herwig} and the jet selection procedure to enhance hadronization effects, based on the \textit{resolved SoftDrop}; results for the charge correlation ratio are discussed in section III, with final conclusions in section IV. Finally, Appendix A discusses more in detail the impact of selecting different parton showers and hadronization models on the present study.

\section{II - Event generation and jet substructure}

DIS events provide a clean environment to study the hadronization mechanism and confinement dynamics, as they produce final states with typically low multiplicities, allowing for more precise measurements. DIS data is also required for the testing and calibration of the theoretical tools, such as MC event generators, used to study the quark-gluon plasma produced in heavy-ion collisions. Here, we build on recent work~\cite{Chien:2021yol} related to electric charge and formation time studies in DIS systems towards refining the sensitivity of novel experimental observables to hadronization effects. As such, we make use of both \textit{Pythia} 8.306 and \textit{Herwig} 7.2 to test a variety of jet observables and selections across the two different phenomenological models - Lund and cluster, respectively. We aim to inform future experimental measurements at the upcoming Electron Ion Collider (EIC), planned to start its activity by the start of the new decade.

As such, we simulate 40 million $ep$ DIS collisions using both MC event generators, anticipated to be 4\% of the total integrated luminosity at EIC~\cite{Chien:2021yol}. We impose the maximum EIC beam energies of $E_e = 18$ GeV and $E_p = 275$ GeV for the electron and proton beams, respectively. A minimum momentum transfer of $Q^2 > 50$ GeV$^2$/c$^2$ is applied on the hard scatterings to ensure the production of high $p_T$ jets and particles are selected with transverse momentum $p_T > 200$ MeV/c to remove high-collinearity to the beam. 

From the list of final-state particles, jet finding via the anti-k$_T$ algorithm~\cite{Cacciari:2008gp} as implemented in \textsc{fastjet} 3.3.0~\cite{Cacciari:2011ma} with a resolution parameter of $R=1$ is performed and inclusive $p_{T, jet} > 7$ GeV/c\footnote{While we closely followed the procedure depicted in~\cite{Chien:2021yol}, we focused on $p_{T,jet} > 7~\rm{GeV/c}$ instead of $p_{T,jet} > 5~\rm{GeV/c}$ to minimize the differences between the hadronization models.} transverse momentum jets are selected within a pseudo-rapidity range of $-3.5 < \eta_{jet} < 1.5$. As we intend to probe hadronization using jets, specifically by studying the electric charges of their highest transverse momentum particles, only jets with at least two final-state charged hadrons are selected. We then perform the SD grooming on jets via the Cambridge/Aachen (C/A) algorithm~\cite{Dokshitzer:1997in} for geometric substructure analysis, since it enforces a strict angular ordering in the clustering trees. Accordingly, the first unclustering step of these trees corresponds to the largest opening angle. For formation time studies, the SD algorithm is applied to $\tau$ reclustered trees~\cite{Apolinario:2020uvt,Apolinario:2024hsm} since the $\tau$ algorithm's distance measure becomes a proxy for the (inverse) formation time of the clustered particles in the high-energy, soft and collinear limits. As such, the first unclustering will generally have the shortest formation time value. 

\begin{figure}[!htb]
  \includegraphics[width=0.5\textwidth]{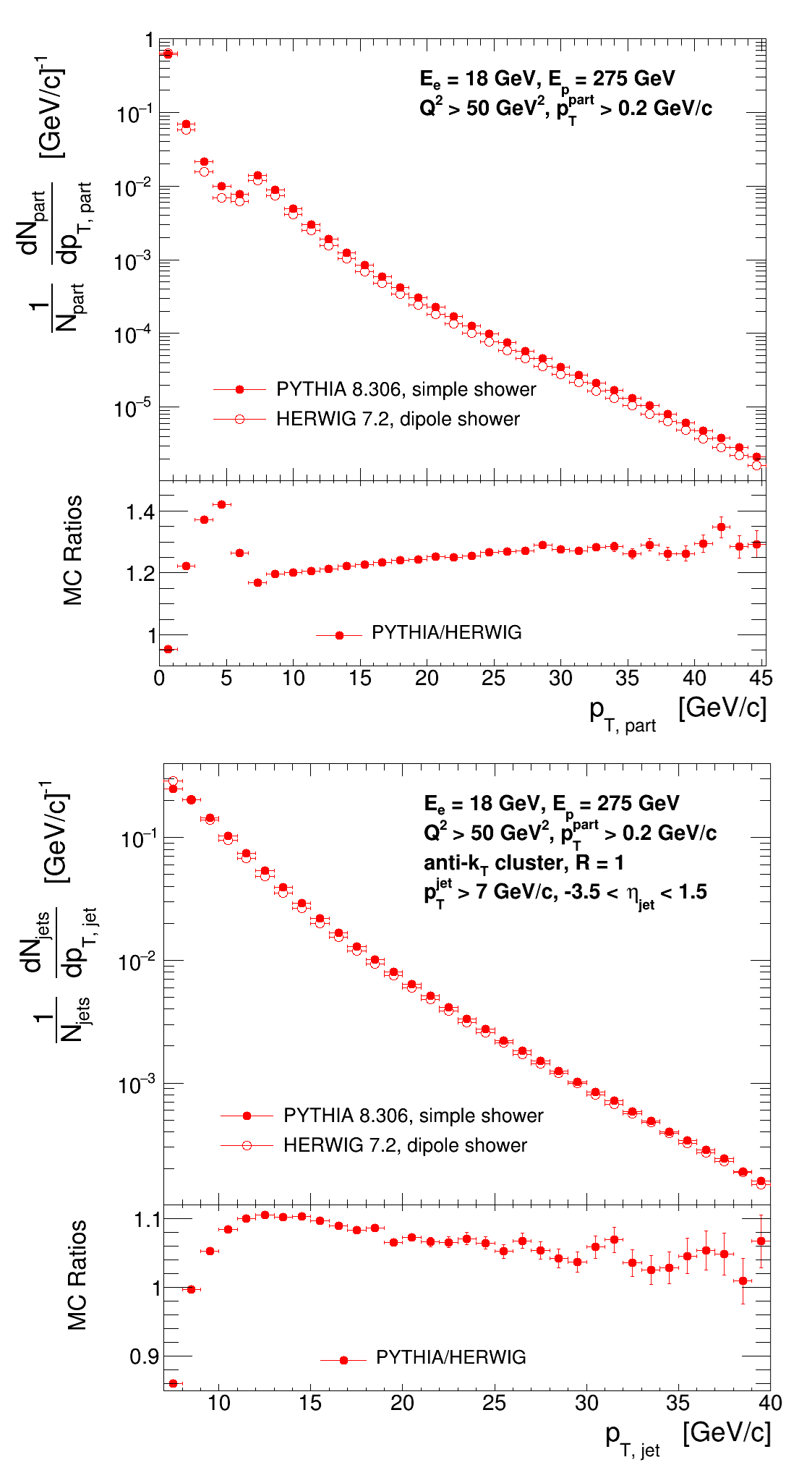}
  \caption{Top panel: particle transverse momentum spectrum for particles with $p_T > 200$ MeV/c for \textit{Pythia}'s simple shower in full red circles and \textit{Herwig}'s dipole shower in open red circles; Bottom panel: jet transverse momentum spectrum with a $p_{T,jet} > 7$ GeV/c selection and $-3.5 < \eta_{jet} < 1.5$ (same colour scheme); both panels have below them the ratios of \textit{Pythia} spectra with respect to \textit{Herwig}.}
  \label{fig:pT}
\end{figure}

With the goal of comparing hadronization models in mind, we select the parton shower descriptions in \textit{Pythia} and \textit{Herwig} that produce the best matches between profiles of the event's particles and jets, with respect to observables like azimuthal angle ($\phi$), pseudorapidity ($\eta$) and transverse momentum ($p_T$). For the chosen collision settings, the results matched the most for \textit{Pythia}'s simple shower and \textit{Herwig}'s dipole shower. For reference, Fig. \ref{fig:pT} shows the particle (top) and jet (bottom) transverse momentum spectra for the DIS events, in full circles for \textit{Pythia} simulations and in open circles for \textit{Herwig} with the settings and selections mentioned above, along with the ratios with respect to \textit{Herwig} results. We can see in the jet spectra that variations between the Monte Carlo generators fall below 10\% for a wide range of transverse momenta.\footnote{Check Appendix A for more details on the effects of the parton shower and hadronization model}.

The momentum fraction at each splitting along the primary branch of the clustering tree is defined as 

\begin{equation}\label{eq:zg}
    z_{g} = \frac{p_{T, 2}}{p_{T, 1} + p_{T, 2}} \quad ,  
\end{equation}
where the transverse momentum of two objects $(1, 2)$ are numbered such that $p_{T, 2} \le p_{T, 1}$. The SD criterion imposed on the C/A or $\tau$ reclustered trees follows the general selection

\begin{equation}\label{eq:SD}
    z_g > z_{cut} \left(\frac{\Delta R_{12}}{R}\right)^{\beta} \quad ,
\end{equation}
where we set the soft threshold $z_{cut} = 0.1$ to remove the divergence at small $z_g$ and the angular exponent $\beta = 0$. The opening angle, commonly referred to as the groomed jet radius or $R_{g}$, is calculated as the distance between the two prongs in the rapidity($y$)-azimuthal($\phi$) angle space $\Delta R$, given by

\begin{equation}\label{eq:Delta_R}
    \Delta R_{12}^2 = \sqrt{ \left(y_1 - y_2\right)^2 + \left(\phi_1 - \phi_2\right)^2 } \quad .
\end{equation}

Finally, we can also evaluate the formation time $\tau_{\text{form}}$~\cite{Apolinario:2020uvt,Apolinario:2024hsm}, which is the time that a quantum state, such as a parton, takes to behave as two independent sources of additional radiation, for example, the two daughter-partons from a splitting. The formation time within jets is associated to varying stages of the parton evolution, ranging from pQCD to npQCD effects. For a jet splitting, $\tau_{\text{form}}$ can be calculated by:

\begin{equation}\label{eq:tau}
    \tau_{\text{form}} = \frac{1}{2 \; E \; z \; (1-z) \; \left(1 - \cos{\theta_{1,2}}\right)} \quad ,
\end{equation}
where $E=E_1+E_2$ is the total energy of the two objects, $\theta_{12}$ the euclidean opening angle between the daughters' 3-momenta and $z$ is the energy fraction of the softer daughter.

The definitions of $z_g$, $R_g$ and $\tau_{form}$ usually apply to the first SD jet clustering step from reclustered trees, but can be also used for any two quantities such as charged particles or other clusterings within the jet, as we will explore in this study. 

\begin{figure}[!htb]
  \includegraphics[width=0.55\textwidth]{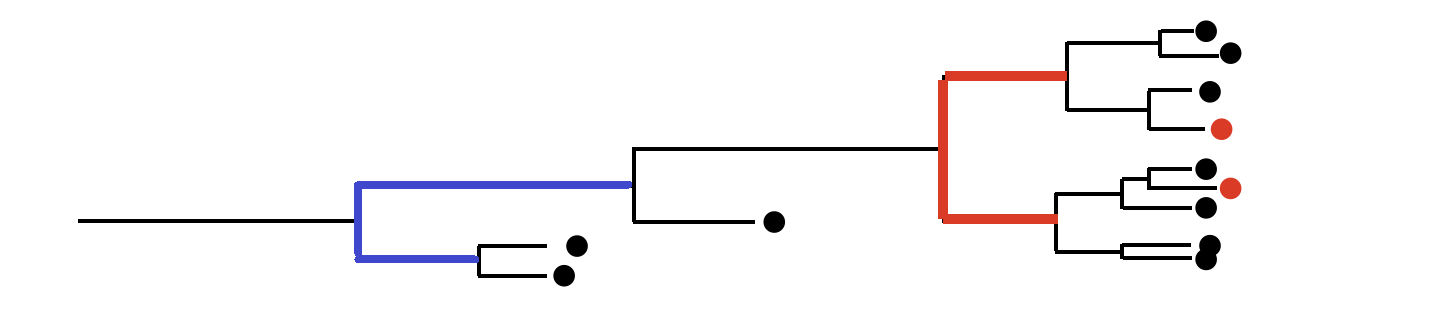}
  \caption{Graphical representation of the clustering tree from a jet and the respective leading charged particles (LCP) in red circles, resolved SoftDrop splitting (RSD) in the red splitting and first SoftDrop splitting (1SD) in the blue splitting.}
  \label{fig:splittings}
\end{figure}

In this paper, we define three distinct \textit{splittings} associated with the jet, diagrammatically represented in Fig. \ref{fig:splittings}. The first unclustering step, starting from the fully clustered C/A jet, that passes our grooming criterion - first SoftDrop splitting (1SD, Fig. \ref{fig:splittings}, blue unclustering) - will be the one we expect to follow a pQCD prescription, as it is the first splitting already shown to yield the Dokshitzer–Gribov–Lipatov–Altarelli–Parisi (DGLAP) splitting functions~\cite{Larkoski:2015lea} from perturbation theory. 

The next we consider are the leading and sub-leading (in terms of their $p_{T}$) charged particles (LCP, Fig. \ref{fig:splittings}, red circles) within the jet. Denoting these two objects as $1, 2$ in Eqs. \eqref{eq:zg}, \eqref{eq:Delta_R} and \eqref{eq:tau} we can identify the parent particle. This is not a realistic splitting in a theoretical sense, since charged particles are produced post hadronization. However, it serves as an useful jet analysis tool, as the LCP is independent of the re-clustering procedure and can be directly measured in experiment. This was previously introduced in~\cite{Chien:2021yol}.

Lastly, as a proxy for this \textit{splitting} within the jet shower, we define the resolved SD splitting (RSD, Fig. \ref{fig:splittings}, red unclustering) which is the first SD unclustering along the primary branch of the jet tree wherein the leading and sub-leading charged particles are separated into their own independent prongs. This clustering step refers to a specific point along the tree where the high-energy, hadronized remnants of the jet are resolved, providing a handle on the depth into the fragmentation pattern at which the two hardest final-state objects from the jet were produced. This can be done via split counting, giving us the so-called RSD depth $N_{RSD}/N_{SD}$, with $N_{RSD}$ accounting for the number of SD splittings along the main jet branch up to and including the RSD, while $N_{SD}$ accounts for the total number of SD splittings in the main branch.

\begin{figure}[!htb]
  \includegraphics[width=0.48\textwidth]{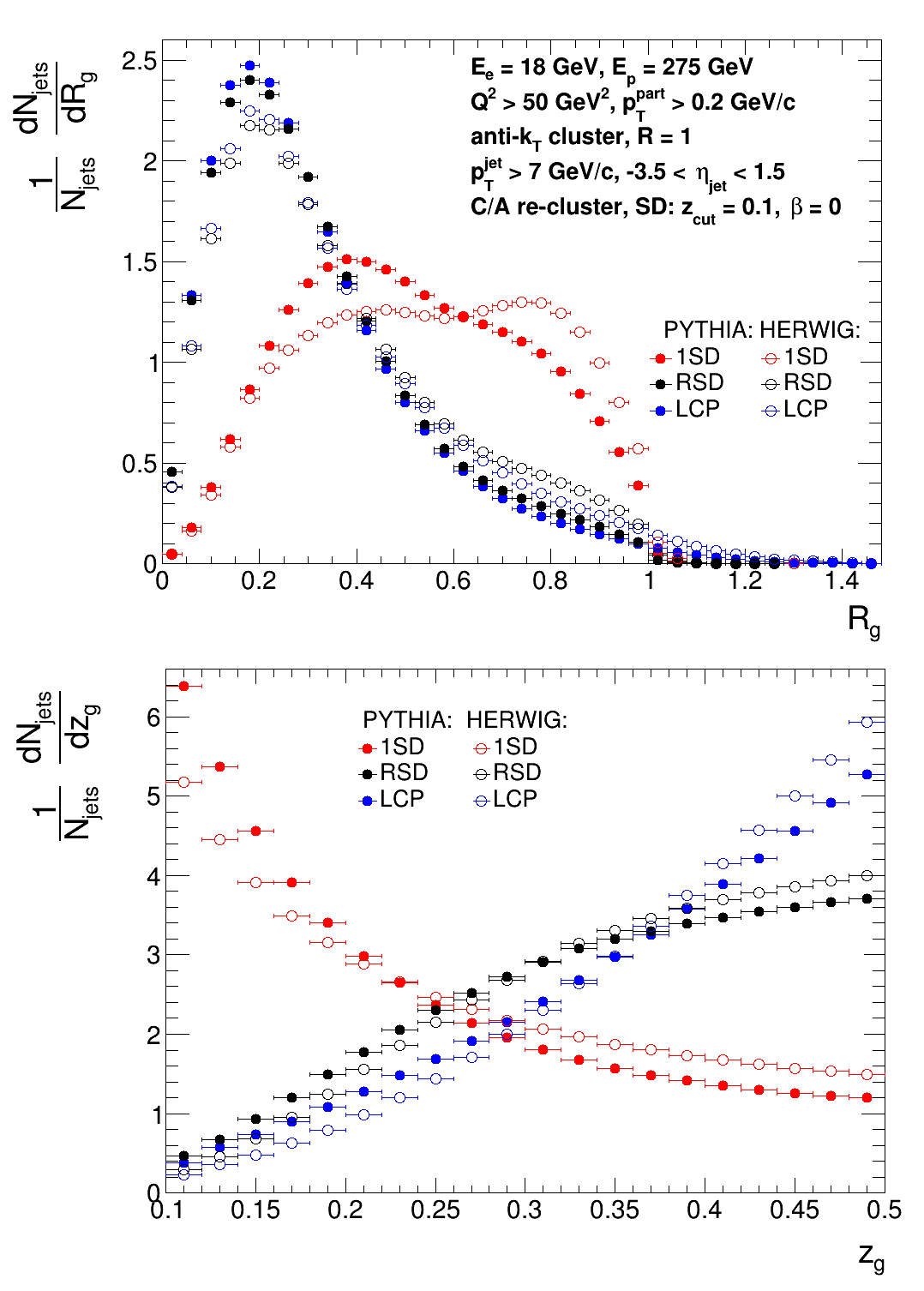}
  \caption{Top panel: groomed opening angle distributions; Bottom panel: groomed momentum fraction distributions; the 1SD, RSD and LCP distributions are self-normalized and represented in red, black and blue markers, respectively, in full circles for \textit{Pythia} jets and open circles for \textit{Herwig} jets.}
  \label{fig:basic}
\end{figure}

Since vacuum parton showers typically generate angular ordered emissions and the re-clustering procedure is performed with an angular ordering variable via the Cambridge/Aachen algorithm, the first unclustering step to survive the SD criterion will produce prongs with larger $R_g$ within the jet (proxy for the opening angle), followed by a sequential reduction in the angle as we travel along the clustering tree. This is reflected on the top panel from Fig. \ref{fig:basic}, where the 1SD, in the blue markers (full circles for \textit{Pythia} simulations and open circles for \textit{Herwig}), will be on average observed at larger angles ($R_{g, 1SD} \approx 0.52/0.56$ for \textit{Pythia}/\textit{Herwig}). Since the RSD, by construction, happens either at or after the 1SD in the clustering tree, their opening angles in the black markers are shown to be smaller ($R_{g, RSD} \approx 0.32/0.36$). As for the LCP, represented via the blue markers, we find that it closely follows the $R_g$ distribution of the RSD splitting, meaning the subsequent branching of the RSD daughters will mostly preserve the opening angle between the leading charged particles ($R_{g, LCP} \approx 0.32/0.36$). The $R_g$ is shown to be sensitive to the hadronization procedure, with significant differences between \textit{Pythia} and \textit{Herwig}.

The pQCD asymmetry in the transverse momentum sharing expected for early splittings (splitting functions highly-peaked at small $z_g$) can also be observed in Fig. \ref{fig:basic}, in the bottom panel, with the 1SD being the most asymmetric splitting, peaking at $z_g = z_{cut} = 0.1$, followed by the RSD and lastly the LCP, both tendentiously symmetrical. The results for both Monte Carlo generators are in good agreement with each other.

\section{III - Formation Time and Charge Correlation Ratio}

\begin{figure}[!htb]
  \includegraphics[width=0.49\textwidth]{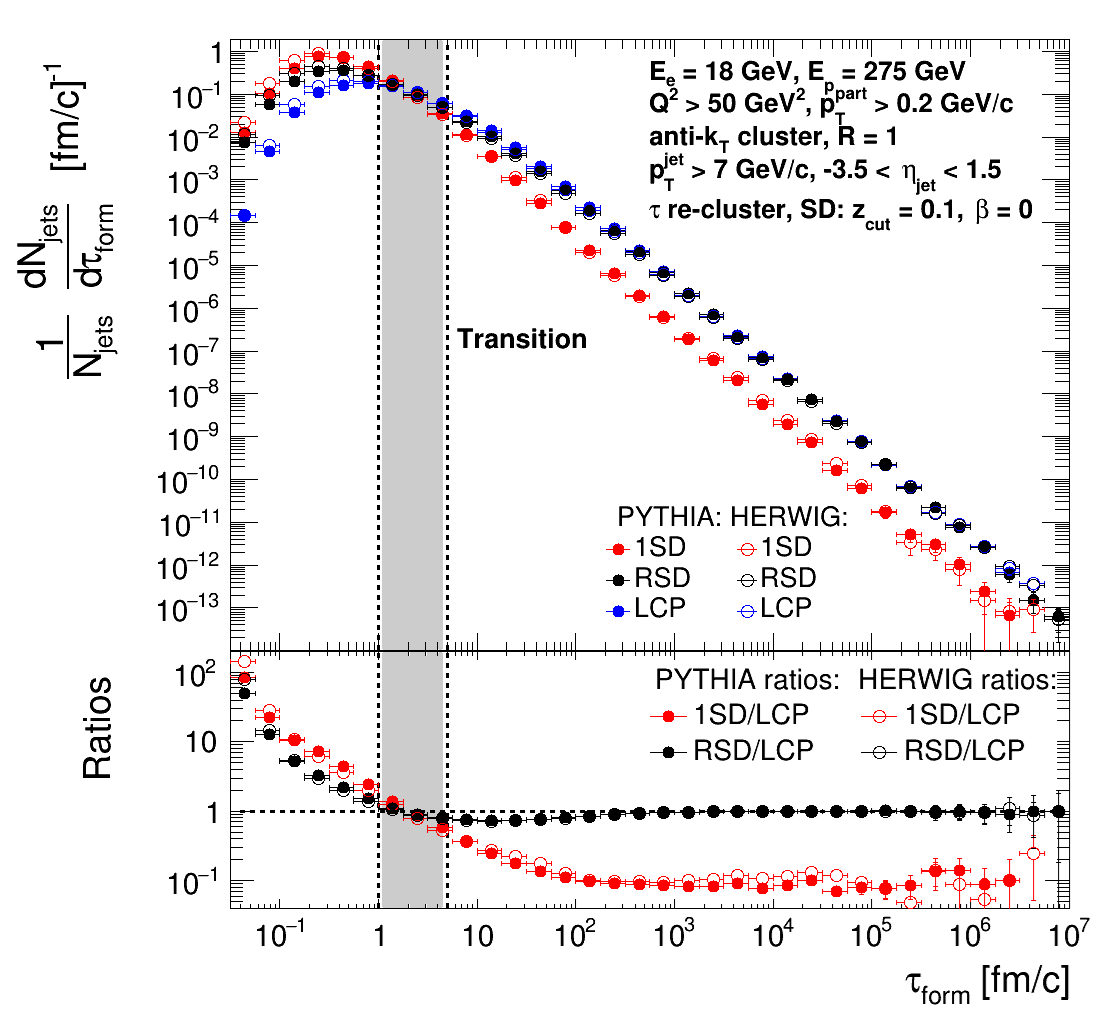}
  \caption{Formation time distributions for the 1SD (red), RSD (black) and LCP (blue) splittings from the sampled jets, where full circles are used for \textit{Pythia} ratios and open circles for \textit{Herwig}.}
  \label{fig:tau}
\end{figure}

The formation time distributions, calculated according to Eq. \eqref{eq:tau} for the 1SD, RSD and LCP splittings from the jets, are shown in Fig. \ref{fig:tau}, along with the ratios with respect to a LCP baseline on the bottom. The 1SD will naturally yield smaller $\tau_{form}$ values as they are selected at the earliest unclustering step, with the largest opening angles and smallest groomed momentum fractions. The LCP $\tau_{form}$ has a characteristic tail at large formation time since the two leading charged particles will predominantly have smaller angles as compared to the jet splittings. We recall that the RSD marks the splitting where the two leading charged particles follow different clustering steps. Interestingly, the RSD distribution is very close to the 1SD for early times ($\tau_{form} \lesssim 1~\rm{fm/c}$, around the proton radius), while its shape is identical to the LCP for later times, as the black plateau in the ratio panel from Fig. \ref{fig:tau} shows. We note that the transition seems to occur around $\tau_{form} \in [1, 5]~\rm{fm/c}$.

With RSD focusing on the high-energy charged particles within the jet, we can directly compare this study with the charge correlation ratio observable introduced here~\cite{Chien:2021yol}. This is defined as the ratio of yields of equally-charged, $d\sigma_{hh}/dX$, and oppositely-charged, $d\sigma_{h\overline{h}}/dX$, leading charged hadrons of flavour $h = \pi^+, K, p$ (pions, kaons and protons), within the jets

\begin{equation}\label{rceq}
    r_c(X) = \frac{d\sigma_{hh}/dX-d\sigma_{h\overline{h}}/dX}{d\sigma_{hh}/dX+d\sigma_{h\overline{h}}/dX} \quad ,
\end{equation}
with respect to a given kinematic variable $X$.

Therefore, $r_c$ acts as a measure of the probability of producing jets with a given LCP charge configuration, with values ranging between $-1 \le r_c \le 1$: positive $r_c$ means a greater probability of jets having both positive or both negative LCP (same sign), while negative $r_c$ signals that jets tend to have both a positive and a negative leading charges (opposite sign). This was recently shown as a direct test of the string breaking phenomenon. 

\begin{figure}[!htb]
  \includegraphics[width=0.48\textwidth]{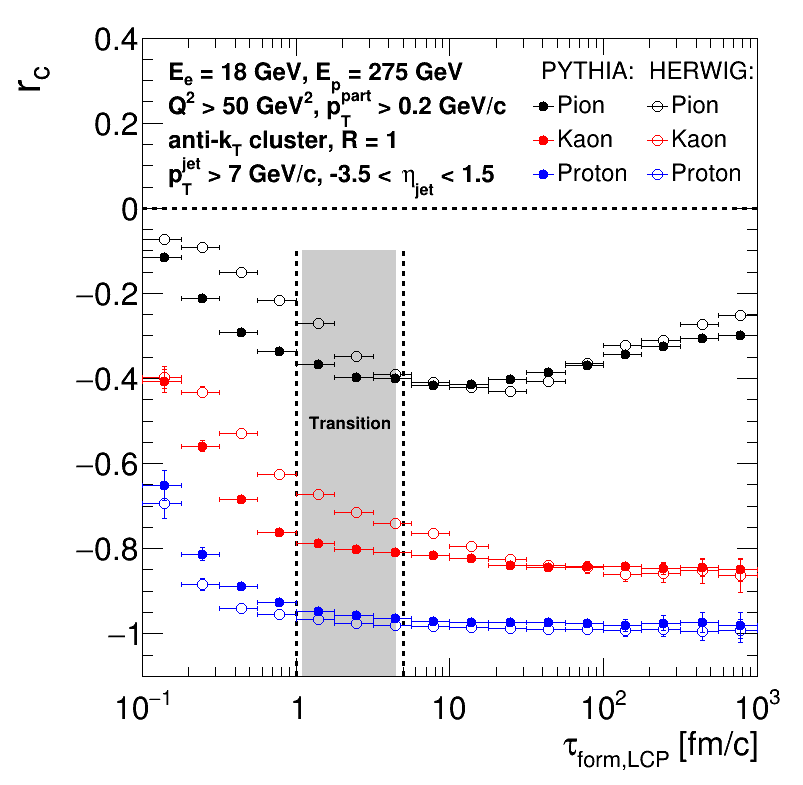}
  \caption{Charge ratios with respect to the LCP formation time for jets with leading charged pions (black), kaons (red) and protons (blue), where full circles are used for \textit{Pythia} ratios and open circles for \textit{Herwig}.}
  \label{fig:rc}
\end{figure}

Since the definition of the $r_c$ is meant to be sensitive to hadronization via the two hardest hadrons, we can study it as a function of the formation time defined in Eq. \eqref{eq:tau}. The charge ratio dependence on the LCP formation time for jets with leading pions, kaons and protons is shown in Fig. \ref{fig:rc} ~\cite{Chien:2021yol}, in the black, red and blue markers, respectively, sampled from jets with at least two identified charged particles. The analysis of \textit{Pythia}(\textit{Herwig}) jets, corresponding to the charge ratios in full(open) circle markers, reveals that 61\%(66\%) have a leading pair of charged pions, 5\%(3\%) have leading kaons and 5\%(6\%) have leading protons, with the remaining jets having mixed-flavour LCP. 

Firstly, the charge ratios are shown to be overall negative, consistently across hadron species. However, the $r_c$ is clearly dependent on the formation time of the LCP. For large $\tau_{\text{form, LCP}}$, the charge ratios are approximately constant at large negative values of $r_c$, particularly for $KK$ and $pp$ pairs within jets. In fact, jets with leading protons have a near 100\% probability of having $p\Bar{p}$, with highly unlikely $pp$ and $\Bar{p}\Bar{p}$ configurations. This reveals a strong preference towards the production of opposite leading charged particles for late LCP jets. We recall that RSD is a good proxy for the LCP for large $\tau_{\text{form, LCP}}$ and that $\tau$ reclustered trees will have the largest formation time values for the latest unclustering steps, i.e., in the last stages of the jet fragmentation pattern. As such, large LCP formation time selects jets whose leading charged hadrons get resolved very late, making them less likely to have subsequent branching randomizing the correlation between the charges.

However, for small $\tau_{\text{form, LCP}}$, the charge ratios become significantly closer to zero, indicating a more equitable probability of having same and opposite sign LCP jets. This result suggests a randomization of the LCP charge profile for early LCP jets, for whom the leading charged particles are resolved early on the jet fragmentation pattern, with several following unclustering steps that smear the charge correlation. In this formation time region, roughly around $\tau_{\text{form, LCP}} \lesssim 5$, \textit{Pythia} noticeably diverges from the \textit{Herwig} predictions for the charge ratio. Therefore, we identify a transition region $\left(1 \lesssim \tau_{\text{form, LCP}} \lesssim 5\right)$ fm/c, that matches the $\tau_{form}$ distribution (Fig.\ref{fig:tau}). The transition region is shaded in grey in both Figs. \ref{fig:tau} and \ref{fig:rc}.

\subsection{Resolved SoftDrop depth}

\begin{figure}[!htb]
  \includegraphics[width=0.48\textwidth]{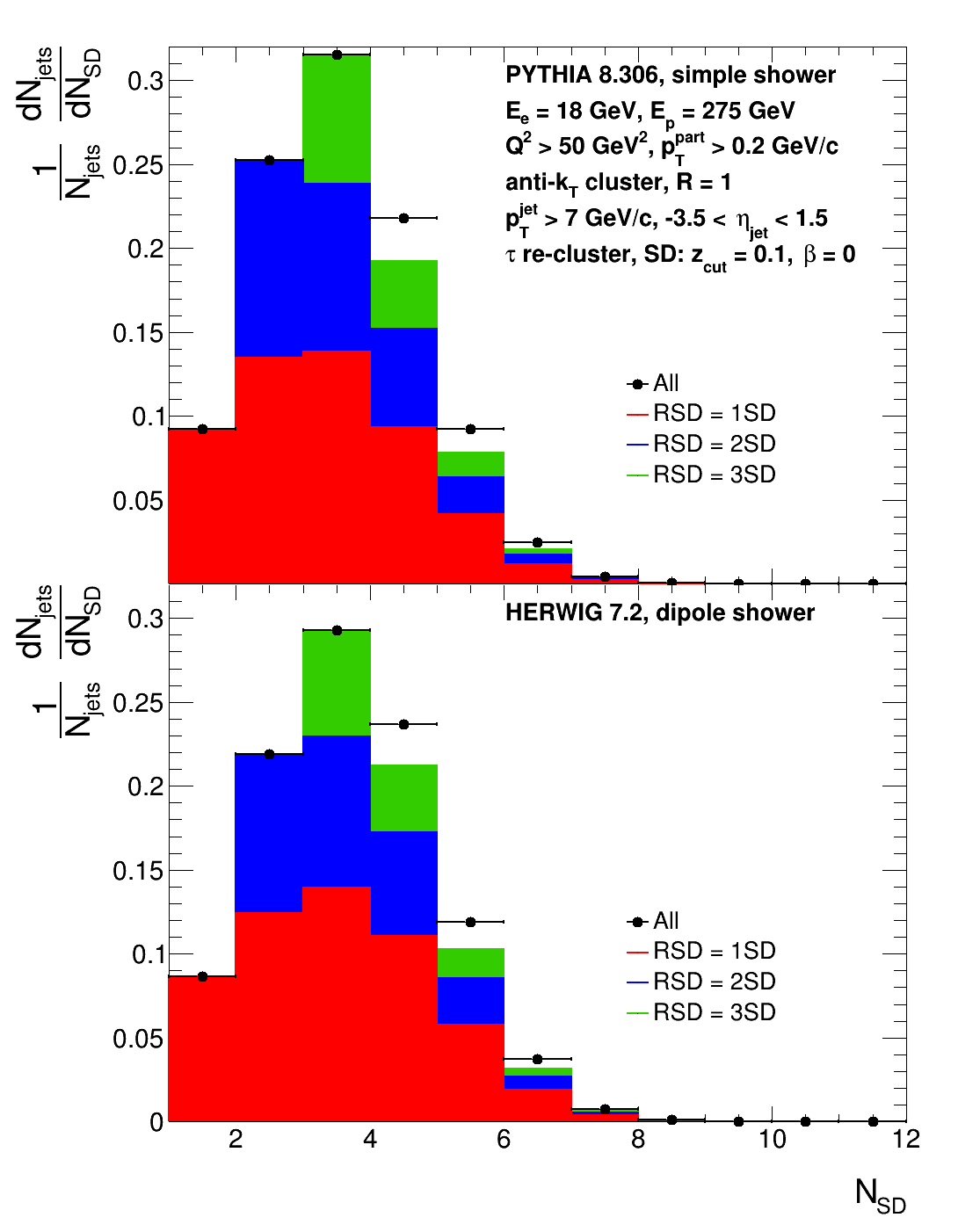}
  \caption{Self-normalized distributions of the total number of SD splittings along the main jet branch (black markers), with the contributions from jets with their first (red), second (blue) and third (green) splittings resolving the leading charged particles (top panel for \textit{Pythia} and bottom panel for \textit{Herwig}).}
  \label{fig:NSD}
\end{figure}

Our goal is to increase the sensitivity of this charge ratio variable to hadronization physics. Namely, jets can have their resolved SD earlier or later in their respective fragmentation pattern. When selecting jets with an identifiable RSD splitting that resolves the leading charged particles from that jet, we observe $\langle N_{SD} \rangle \approx 3$ SoftDrop unclusterings along the main branch for both for \textit{Pythia} and \textit{Herwig}, as can be seen in Fig. \ref{fig:NSD} on the top and bottom panels, respectively. As such, the contributions coming from jets with RSD in the first, second and third SDs are shown as stacked $N_{SD}$ distributions in red, blue and green bars, respectively. We find that 52\% of \textit{Pythia} jets and 55\% of \textit{Herwig} jets have their resolved splitting at the very first SD unclustering of the clustering tree. The RSD will fall on the second SD for an additional 30\%/28\% of jets and on the third SD for 13\%/12\% of them for the kinematic range studied.

\begin{figure}[!htb]
  \includegraphics[width=0.48\textwidth]{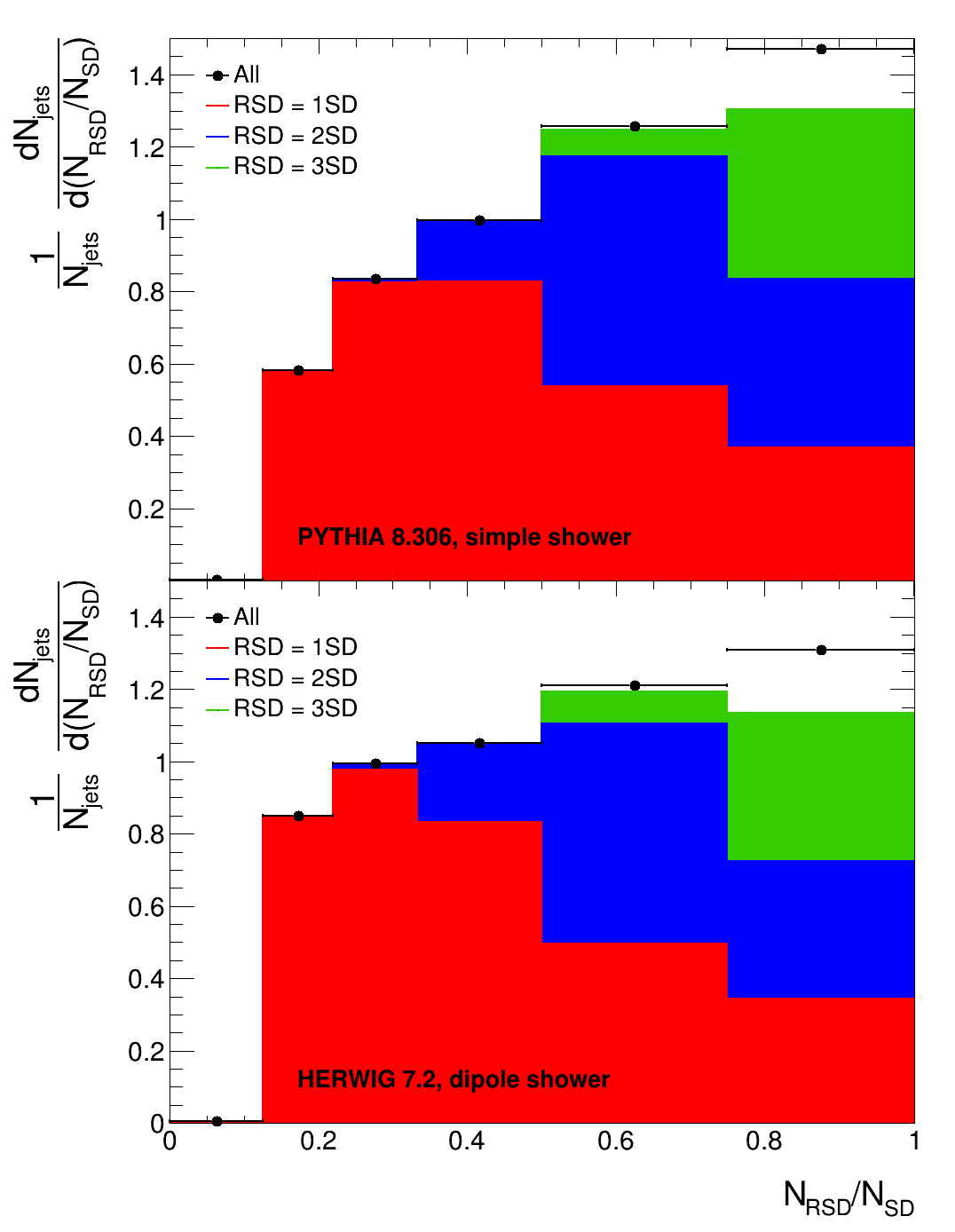}
  \caption{Top panels: self-normalized distributions of the depth of the resolved splitting within the clustering tree (black markers), with the contributions from jets with their first (red), second (blue) and third (green) splittings resolving the leading charged particles (top panel for \textit{Pythia} and bottom panel for \textit{Herwig}).}
  \label{fig:RSDdepth}
\end{figure}

Since jets will have a varying number of total SD splittings, we additionally show the RSD depth $N_{RSD}/N_{SD}$ in Fig. \ref{fig:RSDdepth} with the respective contributions from the location of the RSD along the primary branch (same colour scheme). Jets with an RSD = 1SD that are located at $N_{RSD}/N_{SD} \approx 1$ are jets with only one SD emission along the branch and, although present, are typically rare. Jets with an RSD = 1SD will generally have an early RSD depth, and its distribution will be mostly related to the total number of SD emissions along the primary branch. It is clear that around $N_{RSD}/N_{SD} \approx 0.5$ the population of jets with an early RSD (RSD = 1SD) is no longer the dominant one. As such, we would expect that for this jet population the hadronization effects on the leading charged particles would become more visible. Considering that the charge dipole ratio presents a sensitivity to hadronization effects as well, we proceed to analyze how the $r_c$ can possibly depend on the RSD depth. 

\begin{figure}[!htb]
  \includegraphics[width=.49\textwidth]{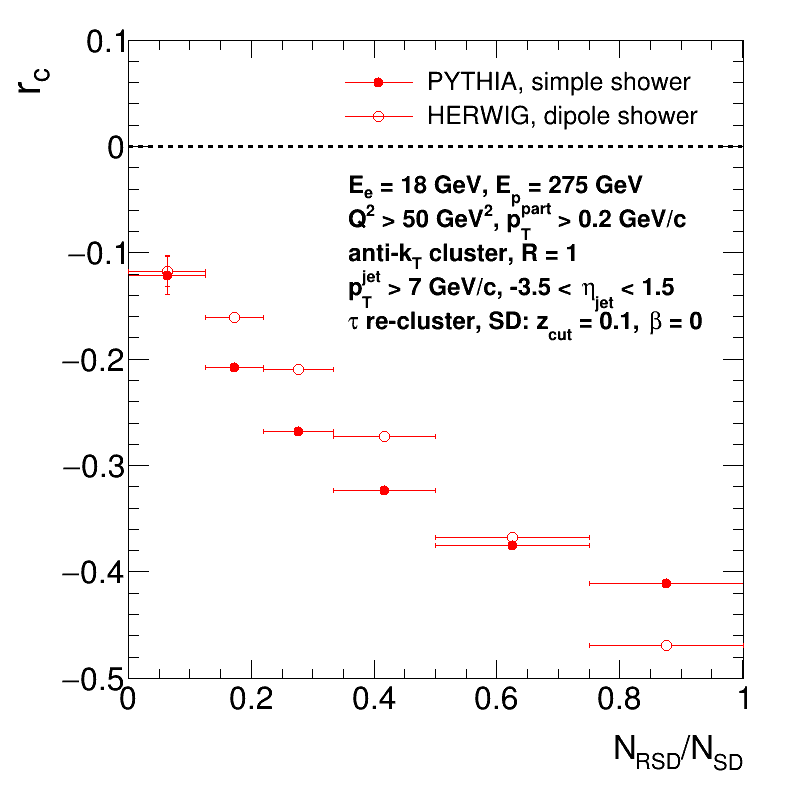}
  \caption{Charge ratios with respect to the RSD depth for jets with any leading hadronic flavours, where full circles are used for \textit{Pythia} ratios and open circles for \textit{Herwig}.}
  \label{fig:rc_RSDdepth}
\end{figure}

As such, for completeness, in Fig.~\ref{fig:rc_RSDdepth}, we further show how the $r_c$ depends on jet substructure topology. Namely, we plot the charge ratio as a function of the RSD depth for both \textit{Pythia} (full circles) and \textit{Herwig} (open circles). For both models, the earlier the RSD takes place, the more random will be the hadronization of the leading charged particles. As the RSD depth increases, the $r_c$ becomes more negative, highlighting that jets with late-stage RSD indeed strongly preserve an opposite charge relation for their LCP.

\subsection{Jet selection based on RSD}

\begin{figure}[!htb]
  \includegraphics[width=0.49\textwidth]{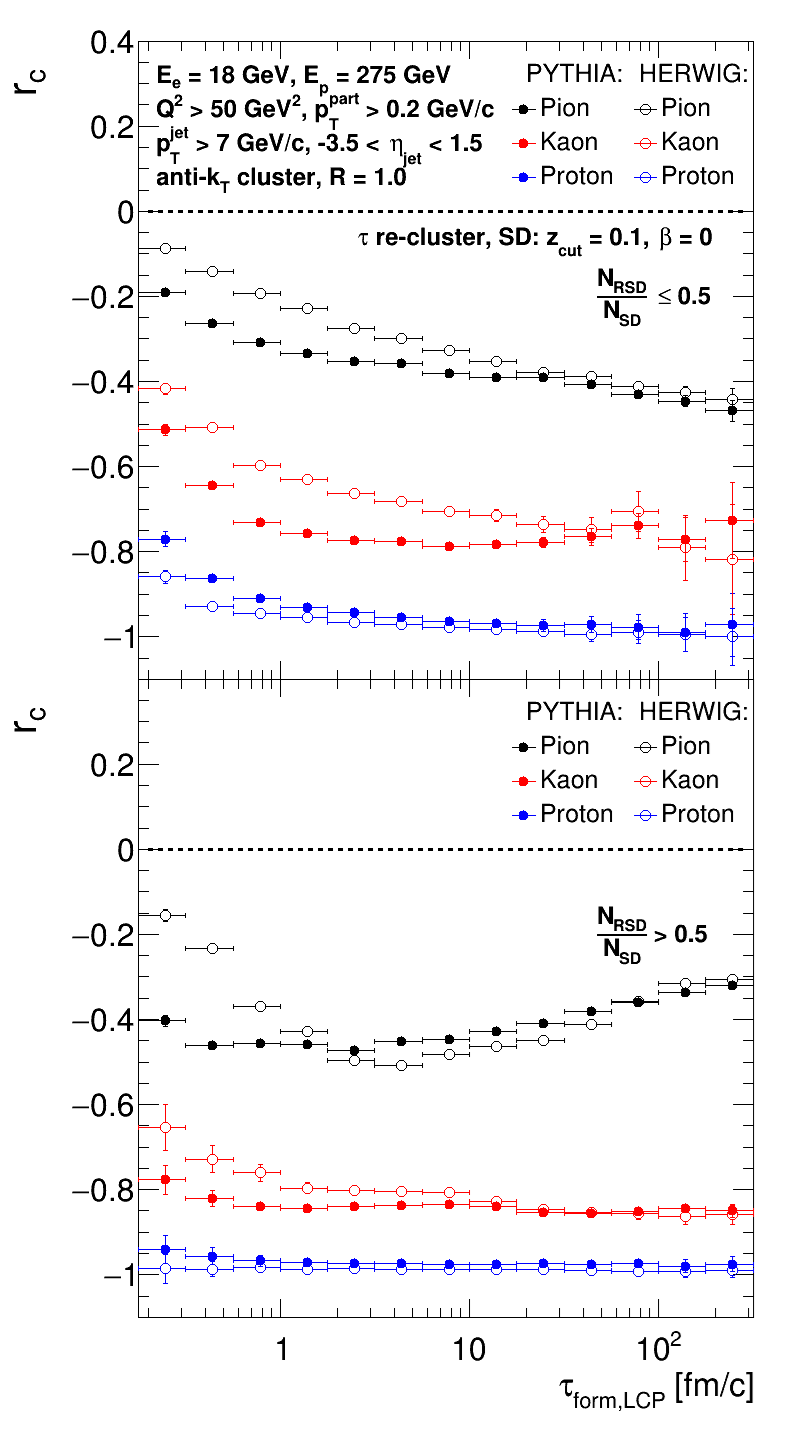}
  \caption{Charge ratios with respect to the LCP formation time for jets with leading pions (black), kaons (red) and protons (blue), where full circles are used for \textit{Pythia} ratios and open circles for \textit{Herwig}; top panel shows early-RSD jets, with small depths; bottom panel shows late-RSD jets, with large depths.}
  \label{fig:rc_RSDdepth_cuts}
\end{figure}

Having clearly identified a dependence of the charge correlation ratio on jet substructure, we will use the additional information provided by the RSD depth to further increase the differences between the hadronization models. Namely, from Fig.~\ref{fig:RSDdepth}, it is possible to identify that jets in which the leading charged particles are coming from clustering steps further along the jet clustering tree will have typically its RSD depth $N_{RSD}/N_{SD} > 0.5$. Using this observation, Fig. \ref{fig:rc_RSDdepth_cuts} replicates the results from Fig. \ref{fig:rc}, but this time selecting separately the jets with early RSD - $N_{RSD}/N_{SD} \leq 0.5$ - shown on the top panel, and jets with late RSD - $N_{RSD}/N_{SD} > 0.5$ - shown on the bottom panel. The $N_{RSD}/N_{SD} \leq 0.5$ jet selection maintains a similar qualitative behaviour between \textit{Pythia}'s and \textit{Herwig}'s $r_c$ dependence on formation time, with an overall strengthening of the LCP opposite sign correlation for later time. However, the $N_{RSD}/N_{SD} > 0.5$ selection showcases significant discrepancies between the generators for small $\tau_{\text{form, LCP}}$ jets. The charge ratios computed with \textit{Pythia} do not have a significant dependence on $\tau_{\text{form, LCP}}$, regardless of the hadronic species of the LCP, leading to approximately constant pion, kaon and proton $r_c$, as evident on the bottom panel of Fig. \ref{fig:rc_RSDdepth_cuts}. However, for the \textit{Herwig} charge ratios, while late time still verifies a plateau-like behaviour, there is a steep climb towards $r_c = 0$ for early-time LCP jets, significantly diverging from \textit{Pythia}'s behaviour.

\begin{figure}[!htb]
  \includegraphics[width=0.49\textwidth]{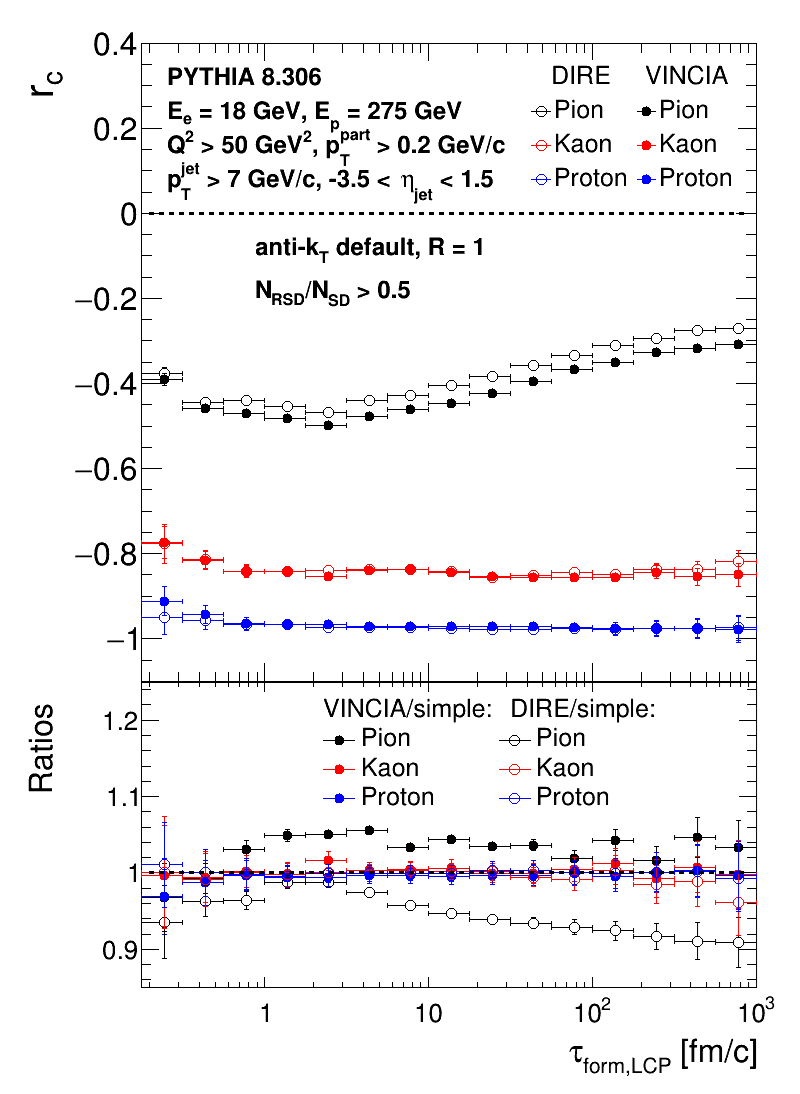}
  \caption{Charge ratio with respect to the LCP formation time for jets with late-RSD and with leading charged pions (black), kaons (red) and protons (blue); full markers belong to jets obtained with the \textit{DIRE} parton shower model and open markers for \textit{VINCIA}, both from \textit{Pythia} 8.306; bottom panel shows the ratio with respect to the $r_c$ from \textit{Pythia}'s simple model.}
  \label{fig:rc_RSDdepth_cuts_pythia}
\end{figure}

\begin{figure}[!htb]
  \includegraphics[width=0.49\textwidth]{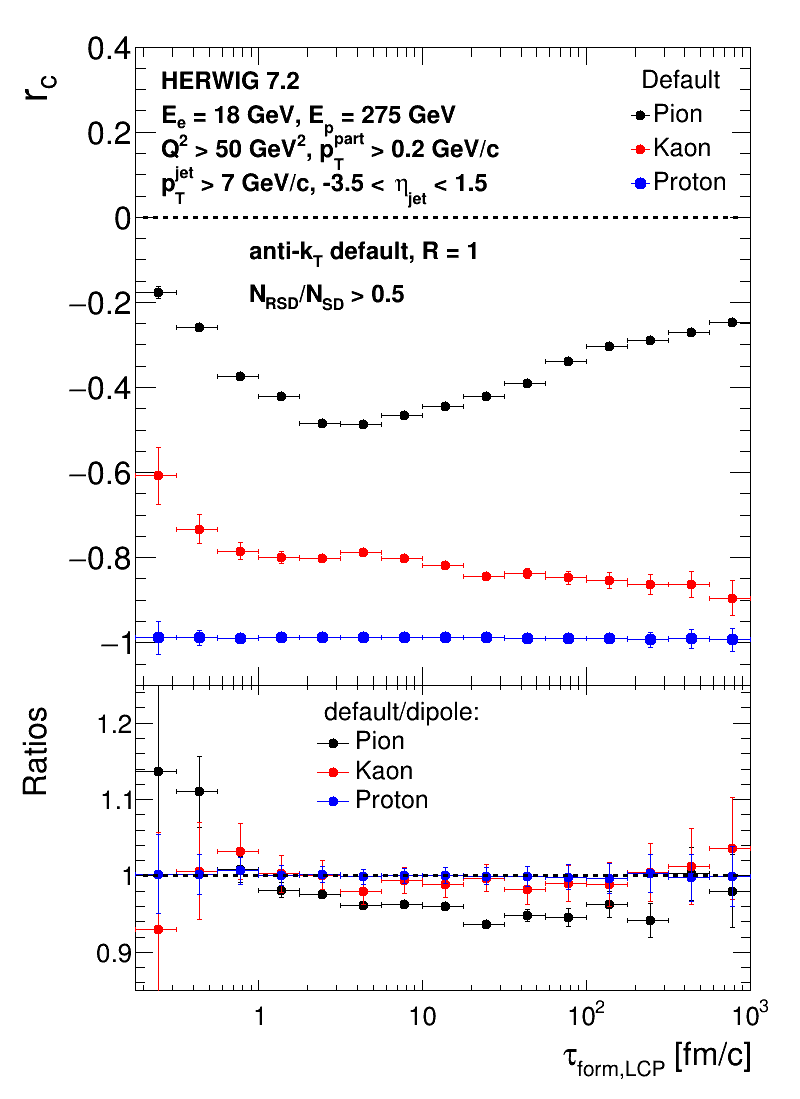}
  \caption{Charge ratio with respect to the LCP formation time for jets with late-RSD and with leading charged pions (black), kaons (red) and protons (blue); results obtained with the \textit{default} parton shower model from \textit{Herwig} 7.2; bottom panel shows the ratio with respect to the $r_c$ from \textit{Herwig}'s dipole model.}
  \label{fig:rc_RSDdepth_cuts_herwig}
\end{figure}

Since we are interested in the charge ratio's sensitivity to hadronization features within varying jet selections, we check the resilience of our findings against primarily changes in pQCD aspects of the models. In Appendix A, we show the charge correlation ratios for a fixed parton shower description, isolating the effects due to Lund string and cluster fragmentation models for hadronization, reproducing the results from Fig. \ref{fig:rc_RSDdepth_cuts} with minor differences. In addition, Figs. \ref{fig:rc_RSDdepth_cuts_pythia} and \ref{fig:rc_RSDdepth_cuts_herwig} quantify the impact of varying  parton shower models on $r_c$. Fig. 10 shows \textit{Pythia}'s charge ratios for $N_{RSD}/N_{SD} > 0.5$, and similarly Fig. 11 the $r_c$ from \textit{Herwig}, with all other parton shower options available in the event generation. We show the \textit{Pythia} simulations using the \textit{DIRE} (full markers) and \textit{VINCIA} (open markers) showers in Fig.~\ref{fig:rc_RSDdepth_cuts_pythia} and Fig. \ref{fig:rc_RSDdepth_cuts_herwig} the $r_c$ from \textit{Herwig} simulations with the \textit{default} shower. The bottom panels in each figure show the ratios with respect to the \textit{simple} and \textit{dipole} baselines we have been using for \textit{Pythia} and \textit{Herwig}, respectively. The plateau nature of \textit{Pythia}'s $r_c$ is shown to be relatively independent of the parton shower model employed, with minimal variations limited to less than 10\% relative to the \textit{simple} shower baseline. In particular, the variations of the $r_c$ for jets with leading kaons (red) and leading protons (blue) are smaller than 2\%. For the \textit{Herwig} charge ratios, we observe that there is only a slight dependence on parton shower models of the early-time $r_c$, where jets with leading charged pions have a slightly less random charge configuration. Similarly, we find jets with leading kaons have a stronger correlation whereas jets with leading protons remain basically the same. We can therefore conclude these charge ratio results to be robust against changes in the perturbative description, isolating the $r_c$ sensitivity to npQCD effects.

\section{V - Conclusions}

Jet substructure tools aimed to reduce non-perturbative contributions unlock a direct comparison between experimental results and analytical calculations. In this manuscript, by excluding the jet population in which non-perturbative contributions are reduced, we identified a jet sample where hadronization models yield qualitatively different results for the charged correlation ratio. By introducing the \textit{resolved SoftDrop splitting}, as the location in the jet clustering tree in which the leading and sub-leading charged particles within a jet are resolved in independent branches, we show that the depth at which this \textit{splitting} is located dictates the overall charge correlation ratio behaviour of the two hadronization models used in this work: Lund strings via \textit{Pythia} and cluster fragmentation via \textit{Herwig}.

We know from Fig. \ref{fig:pT} that the overall event distributions from \textit{Pythia} and \textit{Herwig} are not exactly the same, conflating this comparative exercise with other factors. However, the shear magnitude of the differences observed for the two generators in the late-RSD ($N_{RSD}/N_{SD} > 0.5$), early-LCP ($\tau_{\text{form, LCP}} \lesssim 1$ fm/c) jet selections lead us to conclude the differences observed come predominantly from the different hadronization models - Lund string and cluster fragmentation. This conclusion is reinforced by the robustness of the $r_c$ against varying parton shower descriptions. As such, this selection allows to highlight discrepancies coming from the string model, which retains memory of the incoming parton on the LCP fragmentation, as opposed to the cluster model that lumps the partons in clusters to achieve colour neutrality. In this case, for small $\tau_{form, LCP}$, one expects subsequent unclustering steps to randomize the resulting hadronization pattern.

Therefore, a selection based on the RSD depth reveals a qualitatively different behaviour between the charge ratios predicted by \textit{Pythia} and \textit{Herwig}, magnifying the sensitivity of this variable to the hadronization physics described by the two models.

\section{Acknowledgments}

We thank Isaac Mooney, Laura Havener, Yang-Ting Chien, Youqi Song, Krishna Rajagopal and Jose Guilherme Milhano for their valuable discussions and comments on the manuscript and Andrzej Siodmok and Pratixan Sarmah for their help on accessing state-of-the-art Monte Carlo tools. This work was supported by the European Council (ERC) project ERC-2018-ADG-835105 YoctoLHC; by OE Portugal, Fundação para a Ciência e a Tecnologia (FCT), I.P., project CERN/FISPAR/0032/2021. LA and NOM acknowledges the financial support by FCT under contract 2021.03209.CEECIND and under PhD grant PRT/BD/154611/2022, respectively. RKE would also like to acknowledge funding by the U.S. Department of Energy, Office of Science, Office of Nuclear Physics under grant number DE-SC0024660.

\section{Appendix A: Further studies on the impact of shower and hadronization models}\label{sec:appendix}

\begin{figure}[!htb]
  \includegraphics[width=0.5\textwidth]{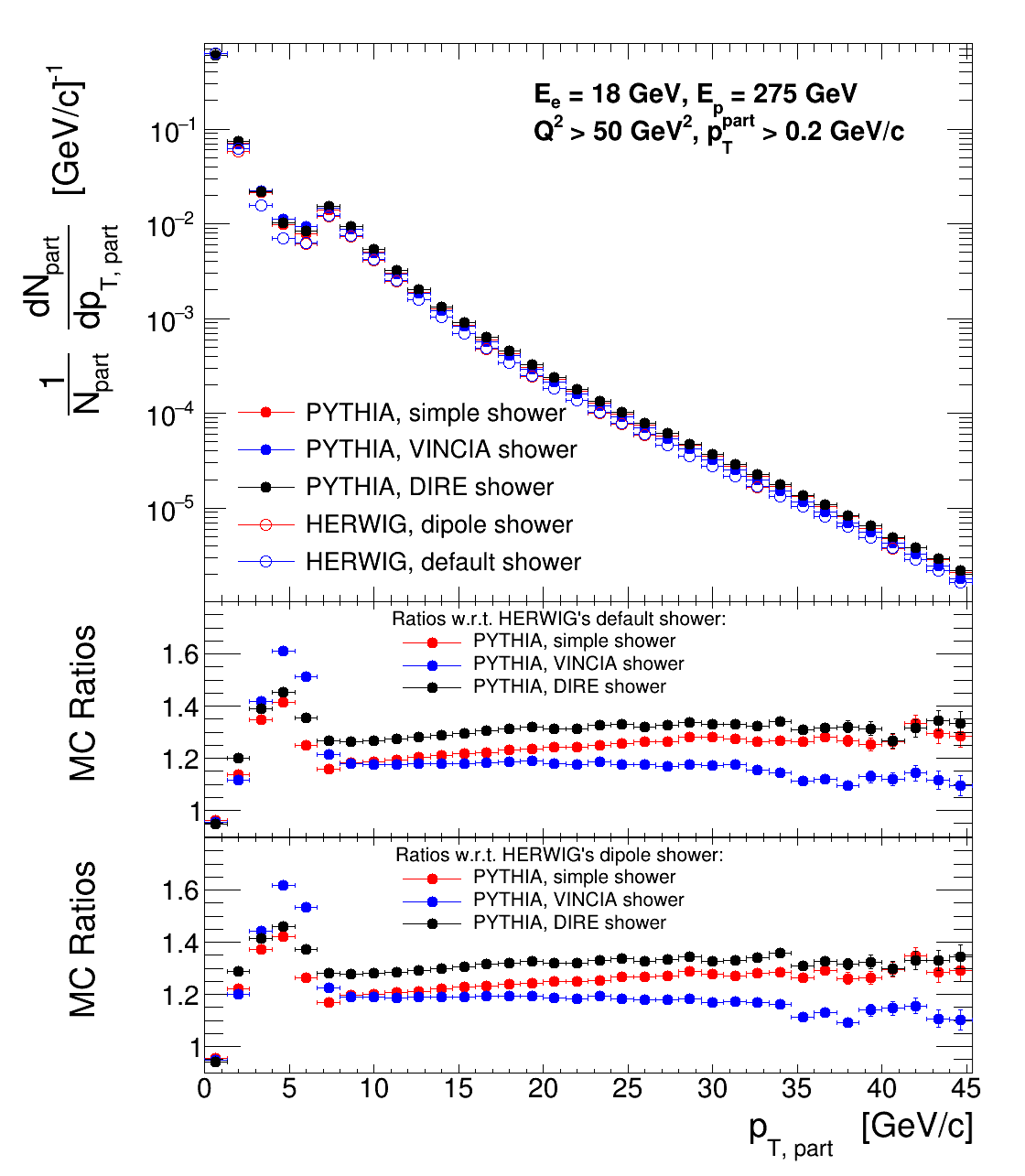}
  \caption{Top panel: particle transverse momentum spectra ($p_T > 200$ MeV/c) for \textit{Pythia}'s \textit{simple} shower (red, full), \textit{VINCIA} (blue, full) and \textit{DIRE} (black, full) and \textit{Herwig}'s \textit{default} shower (blue, open) and \textit{dipole} (red, open); Middle panel: ratios of \textit{Pythia} spectra with respect to \textit{Herwig default}; Bottom panel: ratios of \textit{Pythia} spectra with respect to \textit{Herwig dipole}.}
  \label{fig:pT_part_shower_models}
\end{figure}

Since both Monte Carlo event generators provide more than one parton shower description for the pQCD dynamics, we compared the different options between \textit{Pythia} and \textit{Herwig}. In the top panel of Fig. \ref{fig:pT_part_shower_models}, it is represented the transverse momentum ($p_T$) spectra of final-state particles \textit{Pythia} using Lund string hadronization. The spectra are shown for three different parton showers: \textit{simple} (full red circles), \textit{VINCIA} (full blue circles), and \textit{DIRE} (full black circles). Corresponding data from \textit{Herwig} (cluster hadronization model) using both \textit{default} (open blue circles) and \textit{dipole} (open red circles) showers are also included. The middle and bottom panels display the ratios of \textit{Pythia}’s spectra to those from \textit{Herwig}’s \textit{simple} and \textit{dipole} showers, respectively. We can see that the \textit{simple} shower from \textit{Herwig} most closely aligns with \textit{Pythia}’s output, especially in the dominant low-$p_T$ region. 

\begin{figure}[!htb]
  \includegraphics[width=0.5\textwidth]{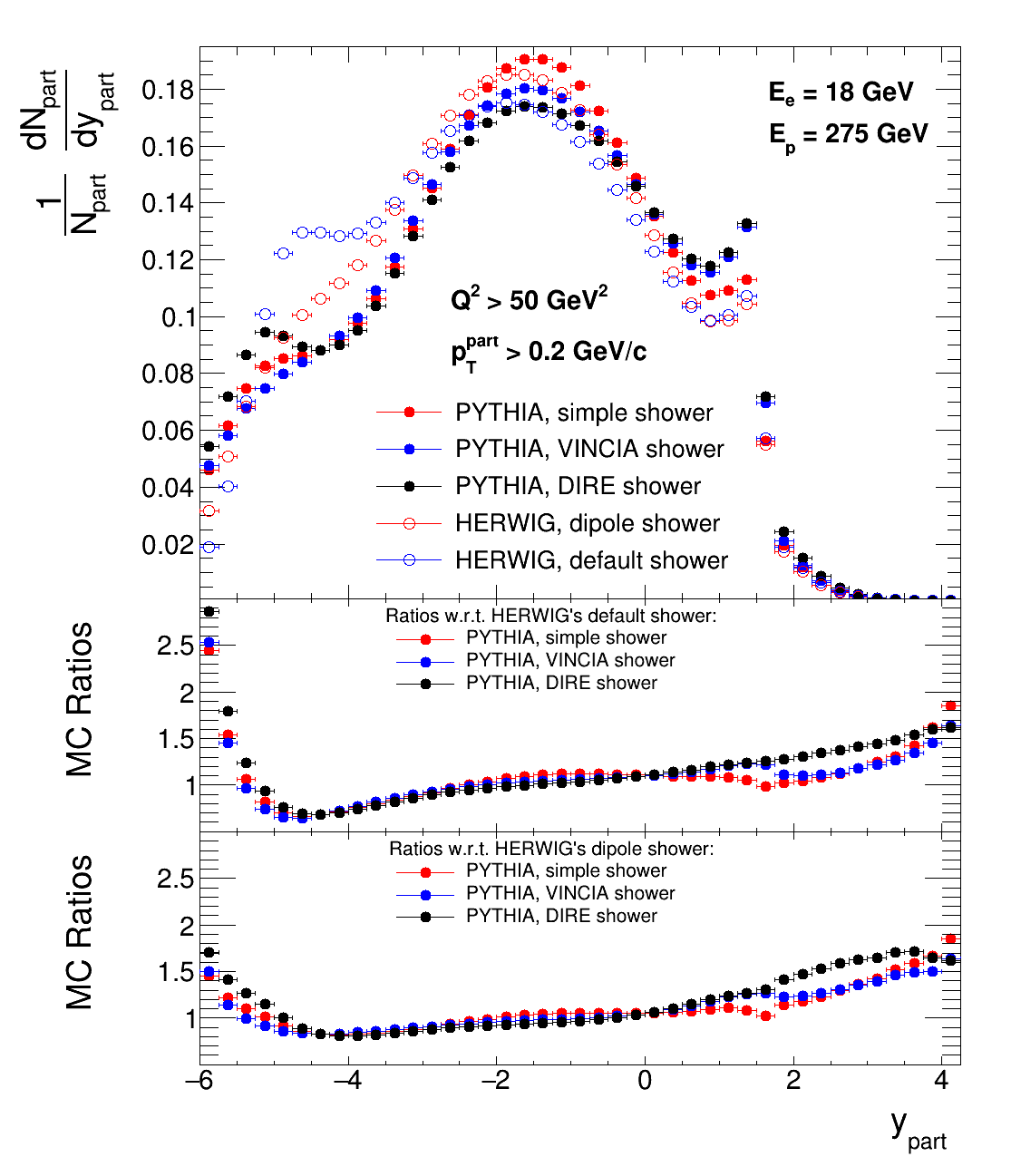}
  \caption{Top panel: particle rapidity for \textit{Pythia}'s \textit{simple} shower (red, full), \textit{VINCIA} (blue, full) and \textit{DIRE} (black, full) and \textit{Herwig}'s \textit{default} shower (blue, open) and \textit{dipole} (red, open); Middle panel: ratios of \textit{Pythia} distributions with respect to \textit{Herwig default}; Bottom panel: ratios of \textit{Pythia} distributions with respect to \textit{Herwig dipole}.}
  \label{fig:rap_part_shower_models}
\end{figure}

The rapidity ($y$) distributions were also assessed and are presented in Fig. \ref{fig:rap_part_shower_models} using the same panel arrangement and marker scheme as previously mentioned. This analysis reinforces the superior alignment of the \textit{dipole} shower from \textit{Herwig} with \textit{Pythia}’s rapidity profiles, thus substantiating our decision to compare DIS simulations using \textit{Pythia}’s \textit{simple} shower and \textit{Herwig}’s \textit{dipole} shower.

\begin{figure}[!htb]
  \includegraphics[width=0.5\textwidth]{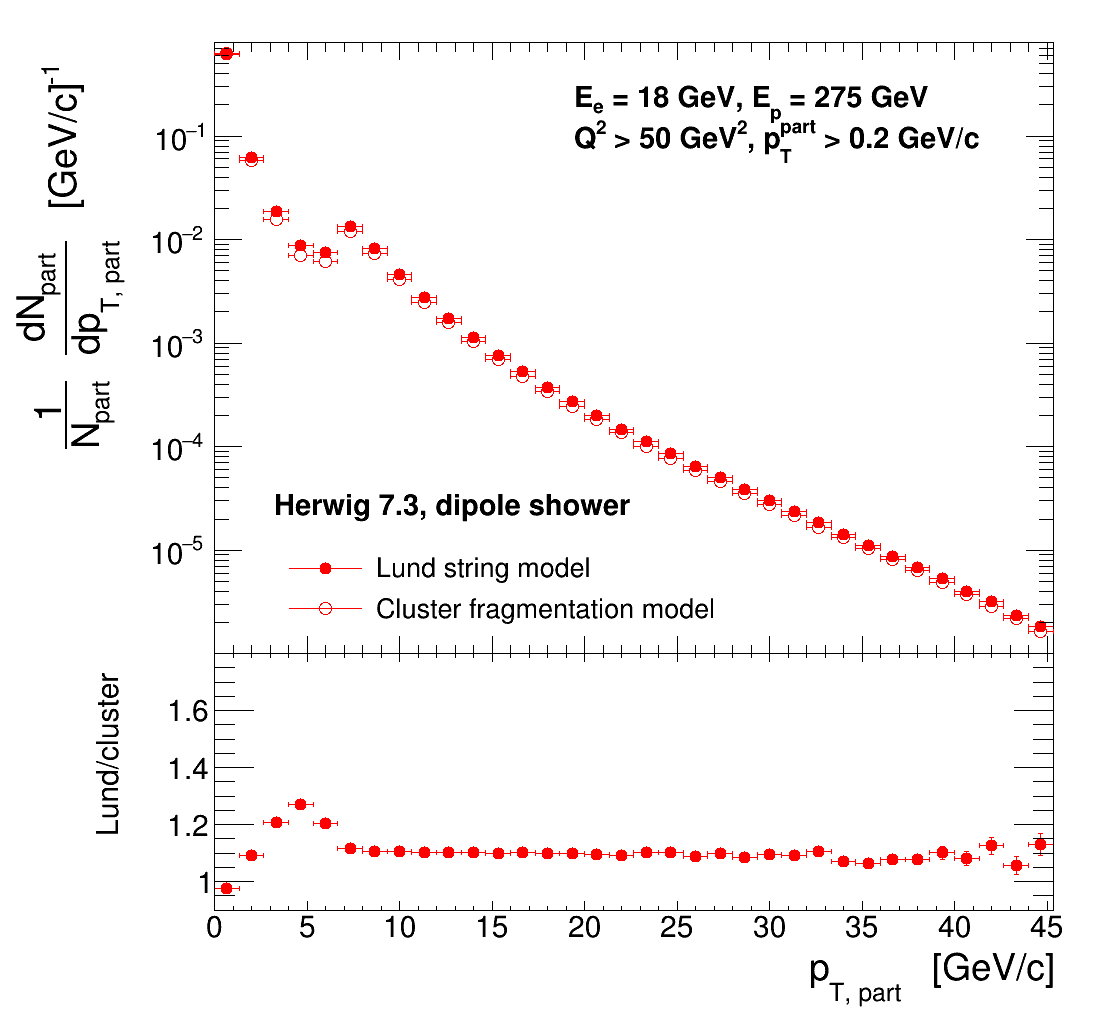}
  \caption{Top panel: particle transverse momentum spectra ($p_T > 200$ MeV/c) for \textit{Pythia}'s Lund string model (red, full) and \textit{Herwig}'s cluster fragmentation model (red, open), both applied to \textit{Herwig}'s \textit{dipole} shower; Bottom panel: ratios of the Lund spectra with respect to the cluster ones.}
  \label{fig:pT_part_hadronization_models}
\end{figure}

\begin{figure}[!htb]
  \includegraphics[width=0.5\textwidth]{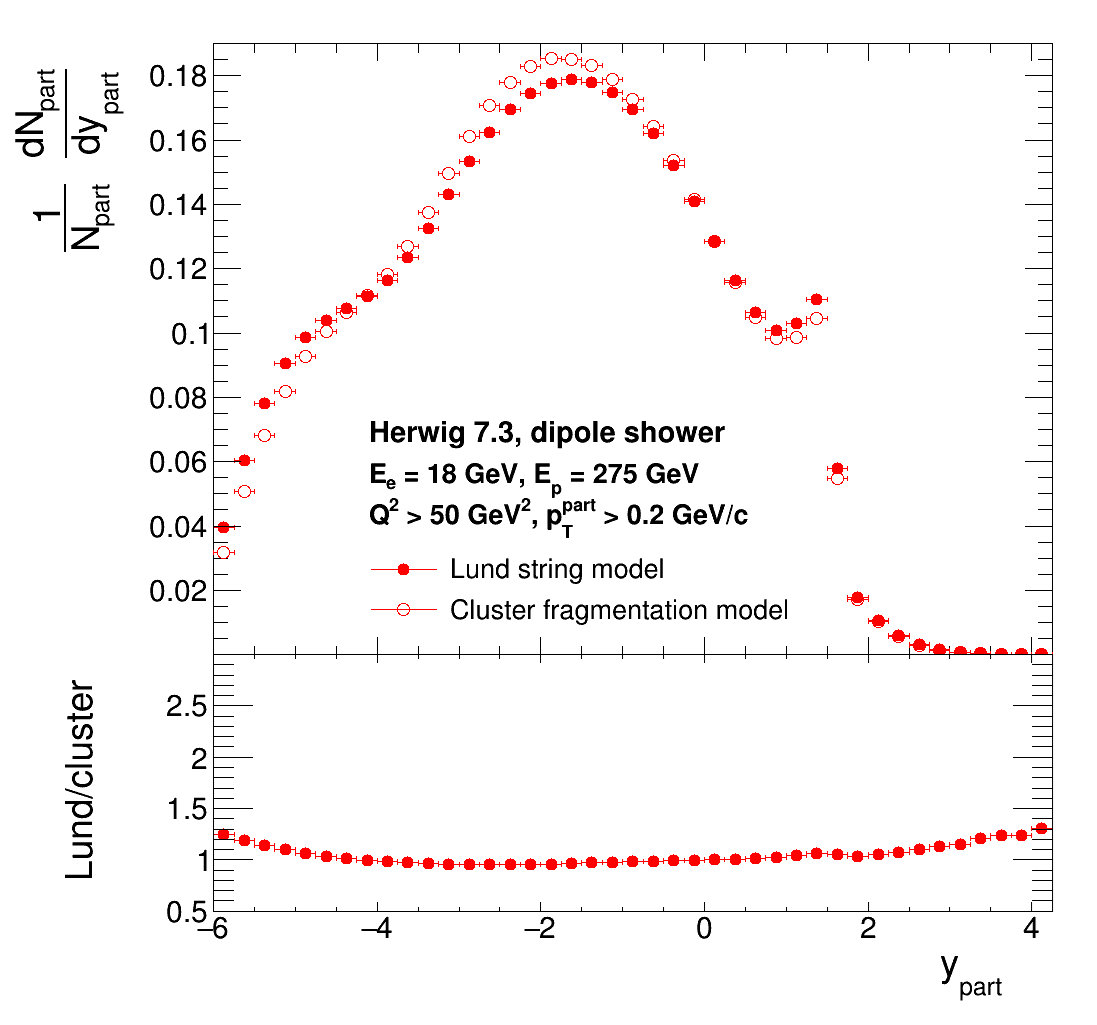}
  \caption{Top panel: particle rapidity for \textit{Pythia}'s Lund string model (red, full) and \textit{Herwig}'s cluster fragmentation model (red, open), both applied to \textit{Herwig}'s \textit{dipole} shower; Bottom panel: ratios of the Lund spectra with respect to the cluster ones.}
  \label{fig:rap_part_hadronization_models}
\end{figure}

Furthermore, we isolated the influence of hadronization models by maintaining a consistent parton shower prescription (angular ordered from \textit{Herwig}) and varying only the hadronization approach. Fig. \ref{fig:pT_part_hadronization_models} shows the $p_T$ spectra for final-state particles using \textit{Pythia}’s Lund string model (full circle markers) and \textit{Herwig}’s cluster model (open circle markers).  This interface with \textit{Pythia} is done via \textit{TheP8I}~\cite{thep8i}, available in the newest \textit{Herwig} 7.3 version~\cite{Bewick:2023tfi}. The rapidity distributions, shown in Fig. \ref{fig:rap_part_hadronization_models}, adhere to the same marker and panel configuration.

\begin{figure}[!htb]
  \includegraphics[width=0.49\textwidth]{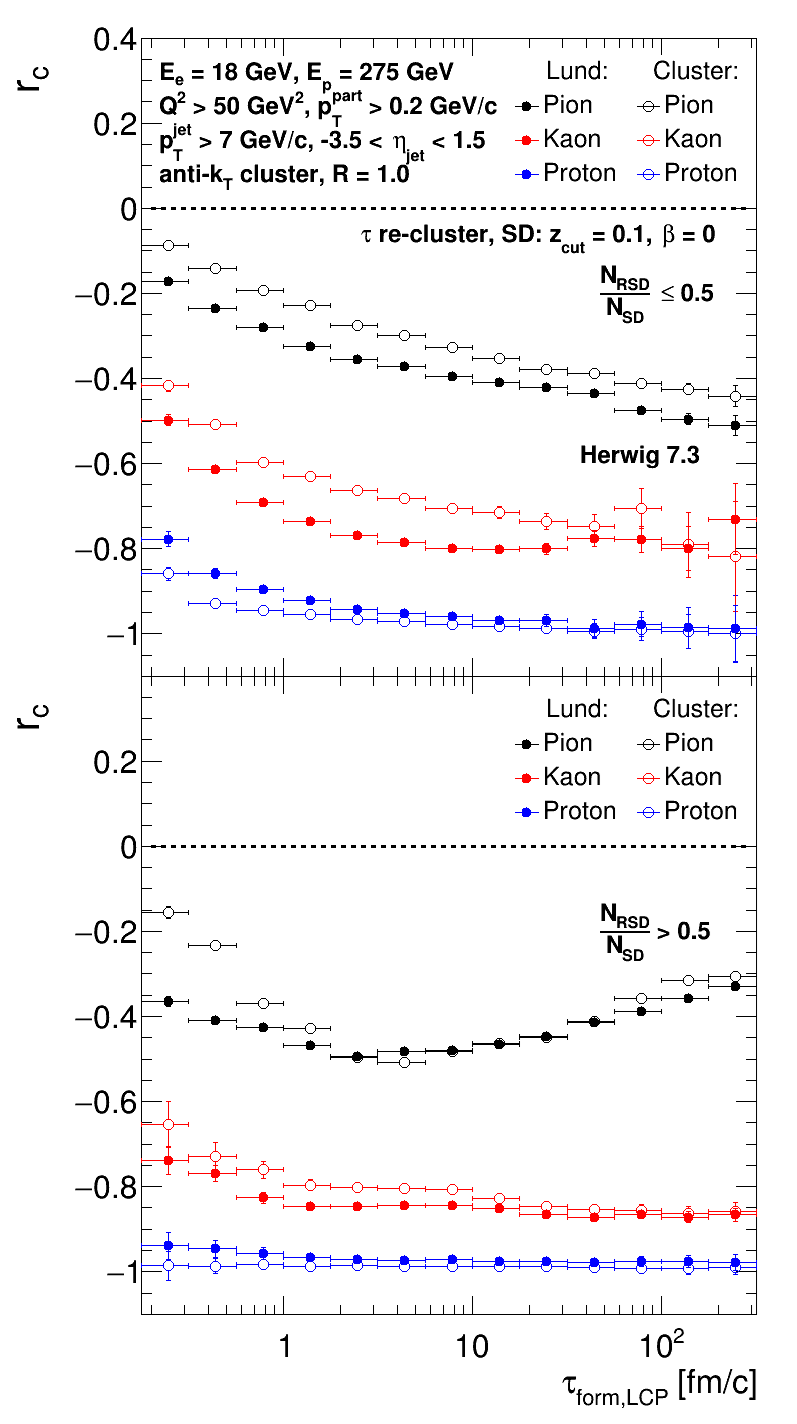}
  \caption{Charge ratios with respect to the LCP formation time for jets with leading charged pions (black), kaons (red) and protons (blue) computed using \textit{Herwig} 7.3, where full circles correspond to the implementation of the Lund string model and the open circles to the cluster fragmentation model; top panel shows early RSD jets and bottom panel shows late RSD jets.}
  \label{fig:rc_cuts_a}
\end{figure}

Given the small differences on the particle and jet yield differences with respect to the main text results, we aimed to further test the resilience of the qualitative discrepancies between the charge ratios from \textit{Pythia} and \textit{Herwig} for late RSD jets (see Fig. \ref{fig:rc_RSDdepth_cuts}) against the perturbative description of the parton shower. As such, Fig. \ref{fig:rc_cuts_a} shows the equivalent charge ratios using the \textit{Herwig} angular ordered parton shower description, but with the full circle markers representing the Lund string model as hadronization mechanism, and the open circle markers the cluster model.

\begin{figure}[!htb]
  \includegraphics[width=0.49\textwidth]{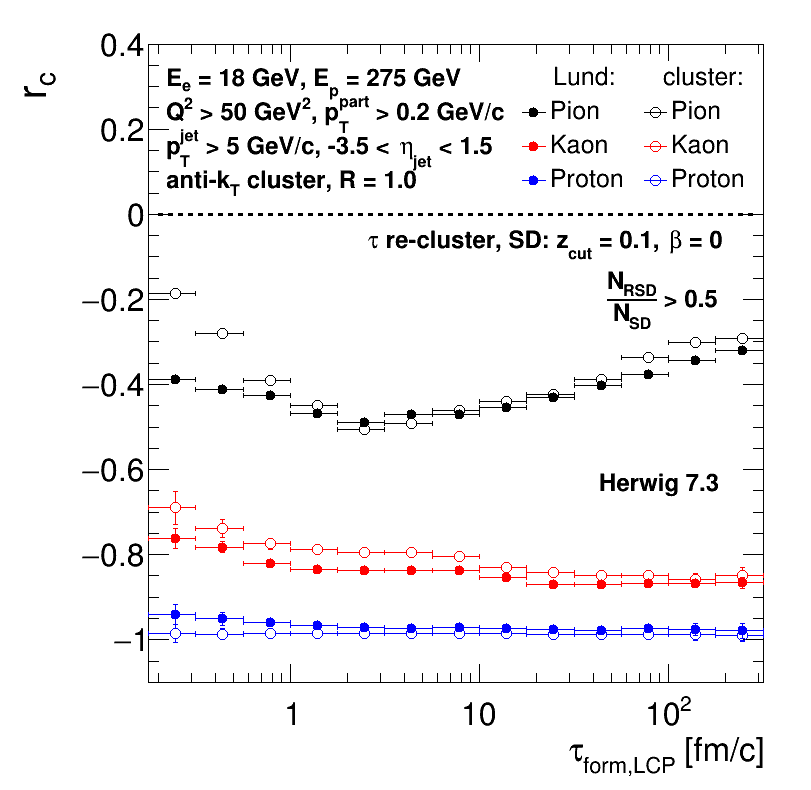}
  \caption{Charge ratios with respect to the LCP formation time for late RSD, $p_{T,jet} > 5$ GeV/c jets with leading charged pions (black), kaons (red) and protons (blue) computed using \textit{Herwig} 7.3, where full circles correspond to the implementation of the Lund string model and the open circles to the cluster fragmentation model.}
  \label{fig:rc_cuts_pT5}
\end{figure}

This assessment continues to highlight a significant divergence in charge ratios for early-time LCP and late RSD jet selections, despite mild numerical differences. Owing to a generally more consistent agreement between the event profiles generated by both models compared to the initial \textit{Herwig} versus \textit{Pythia} comparison, we expanded this analysis to include jets with $p_{T,jet} > 5$ GeV/c, as suggested in \cite{Chien:2021yol}. The outcomes for the lower momentum jets are shown in Fig. \ref{fig:rc_cuts_pT5}, focusing on late RSD jets, which better align with the findings presented in the bottom panel of Fig. \ref{fig:rc_RSDdepth_cuts}.

\bibliography{refs.bib}

\end{document}